\documentclass[10pt,twocolumn]{article}

\usepackage[a4paper,margin=1.8cm]{geometry}

\usepackage{amsmath,amssymb,bm}

\usepackage{graphicx}
\usepackage{booktabs}
\usepackage{multirow}

\usepackage[T1]{fontenc}
\usepackage{lmodern}

\usepackage[colorlinks=true,
            linkcolor=blue,
            citecolor=blue,
            urlcolor=blue]{hyperref}

\usepackage{cleveref}

\usepackage[font=small,labelfont=bf]{caption}

\usepackage{authblk}

\title{\textbf{Plasmonic Bi-Cavity Nanostructure for Efficient Light Collection and Localization}}

\author[1,2,3]{Vitor Monken}
\author[2]{Raul Correa}
\author[3]{Hudson Miranda}
\author[3]{Cassiano Rabelo}
\author[4]{Rafael Nadas}
\author[5]{Thiago L. Vasconcelos}
\author[2]{Luiz Gustavo Cancado}
\author[1,2,*]{Ado Jorio}

\affil[1]{Programa de Pós-Graduação em Inovação Tecnológica e Propriedade Intelectual,Universidade Federal de Minas Gerais, Belo Horizonte, Minas Gerais 30123-970, Brazil}
\affil[2]{Departamento de Física, Universidade Federal de Minas Gerais, Belo Horizonte, Minas Gerais 30123-970, Brazil}
\affil[2]{FabNS, Belo Horizonte, Minas Gerais 31310-260, Brazil}
\affil[2]{Institut fur Physik, Humboldt-Universitat zu Berlin, 12489 Berlin, Germany}
\affil[2]{Divisao de Metrologia de Materiais, Inmetro,Duque de Caxias, Rio de Janeiro 25250-020, Brazil}
\affil[*]{E-mail: adojorio@fisica.ufmg.br}

\date{\today}

\begin{document}

\twocolumn[
\maketitle
\begin{abstract}
\noindent
Tip-enhanced Raman spectroscopy (TERS) typically relies on high-NA excitation to generate a strong axial field at the tip apex, which shortens the working distance and constrains sample geometries. We show that a plasmonic bi-cavity tip, the plasmon-tunable tip pyramid (PTTP), co-tuned in nanopyramid length \(L\) and plateau length \(W\) supports a hybrid antenna--cavity mode that funnels energy to the apex under radially polarized, on-axis excitation, even with a dry objective of \(\mathrm{NA}=0.75\). Finite-element simulations identify \(W\) as a design-critical parameter that sets an in-plane surface-plasmon-polariton (SPP) Fabry--P\'erot-like resonance; co-tuning \((L,W)\) yields a periodic series of maximal apex \(\lvert E\rvert^{2}\). Experiments on monolayer graphene confirm near-field enhancement and reproduce the characteristic annular TERS point-spread function (PSF) with \(\mathrm{NA}=0.75\). Relaxing the NA requirement increases working distance and compatibility with constrained environments, pointing to practical, deployment-ready nano-Raman instrumentation.
\end{abstract}
\vspace{0.5cm}
]

\section{Introduction}
Tip-Enhanced Raman Spectroscopy (TERS) combines the high spatial resolution of scanning probe microscopy (SPM) with the chemical sensitivity of Raman spectroscopy, allowing detailed nanoscale analysis of chemical, structural, and electronic properties \cite{Hayazawa2000, Stockle2000, Anderson2000, Sonntag2014, Shi2017, Hoppener2024, Li2024, Jorio2024}. The widespread adoption of the TERS technique depends on reproducible probes that deliver consistent performance  \cite{novotny2012principles}. These probes must efficiently couple with incoming far-field (FF) excitation light and emit the scattered light back into the FF regime. Additionally, the tip must channel the received energy towards the tip apex, exciting the sample locally within the nanometer scale, i.e. in the near-field (NF), and to properly collect the scattered field generated by the sample in the NF regime. Several researchers worked to develop this field \cite{Foti2022,Lu2018,Calafiore2017,grayholes,Caselli2015,Bao2013,Lindquist2013,Novotny2011, McCourt2024,Meng2023,Slekiene2023}, and here we would like to specifically cite the relevant contributions from professor Renato Zenobi, both in the development of TERS probe designs \cite{Zenobi-1,Zenobi-2,Zenobi-3,Zenobi-4,Zenobi-5} and applications\cite{Human-cel,phaseseparation,Zheng2019,Opilik2015,Stadler2011}. To achieve these goals, a well-established architecture \cite{Achim2003} uses radially polarized light excitation propagating along the dipole axes, focused in the tip apex with a high numerical aperture (NA) objective lens \cite{Novotny1997, Achim}. To ensure axial polarization of the excitation at the tip, the radially polarized beam has to pass through the substrate and the sample in a bottom on-axis illumination. However, the necessity of high-NA optics is undesirable because it imposes limits to the distance between the focusing lens and tip, and therefore on the type of sample to be studied. 

We show that the plasmonic behavior of a well-established TERS probe composed of two connected parts that, together, best meet the efficiency requirement for both the far-field and near-field, considering radially polarized light propagating along the tip axes, without the need of high numerical aperture: a nanopyramid atop a plateau, named plasmon tunable tip pyramid (PTTP) \cite{Vasconcelos2018}. Previous studies focused on optimizing the nanopyramid's length (L) to provide the best possible plasmonic cavity for a given excitation energy.\cite{Vasconcelos2018,Oliveira2020,Miranda2020}. Here, using  finite element method (FEM), we show that the plateau length (W) is also a critical parameter in optimizing the NF enhancement at the tip's apex. This parameter modulates the plasmonic resonance, influencing the enhancement factor and enabling either amplification or suppression of the apex near-field. By combining experimental and simulation results, this study provides new insights into the design of TERS probes, enhancing our understanding of how structural parameters influence plasmonic behavior and opening new avenues for optimizing tip-enhanced nanophotonic systems.

\section{Methods}
\subsection{Numerical simulations}
\begin{figure}[htbp]
\centering
\includegraphics[width=8.4cm]{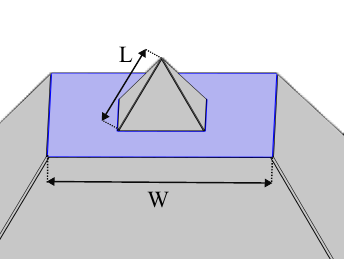}
\caption{Computational model of the nanostructure highlighting the plateau and the optimization parameters. Being L the nanopyramid length and W the plateau length.}
\label{pttp}
\end{figure}

To accurately model the excitation conditions relevant to bottom-illumination TERS, we developed a customized background field expression representing a radially polarized laser beam tightly focused by a low-NA objective lens. This was achieved by discretizing the focal field profile using the method described by \cite{Marrocco2009}, enabling the accurate representation of longitudinal and radial field components in the simulation domain. A Python-based routine was implemented to generate field expressions parametrized by wavelength, beam waist, focal length, numerical aperture (NA), and refractive index. These vector fields were implemented in COMSOL Multiphysics using the Wave Optics module in the frequency domain. The plasmonic nanostructure geometry equivalent to a plasmon-tunable tip pyramid (PTTP) \cite{Vasconcelos2018} was constructed and parametrized in COMSOL, where we performed parametric sweeps of the two key dimensions: the nanopyramid length (L) and the plateau length (W), as shown in figure \ref{pttp}. Both L and W are controllable parameters during probe manufacturing as shown in \cite{Oliveira2020}.

\subsection{Experimental setup and procedures}
For the experimental work, we use a Porto prototype system \cite{Rabelo2019}. It consists of a backscattering micro-Raman system specially designed for TERS, coupled with a shear-force tuning-fork based SPM scan head. The scan head features piezo-driven positioning in X, Y, and Z axes, allowing precise alignment of the probe with the laser focal spot. Coarse positioning is done with piezo stick-slip actuators, while focus control is handled by a separate closed-loop piezo actuator. Sample scanning is achieved with a nanopositioning stage providing sub-nanometric accuracy. A 632.8 nm He-Ne laser (20 mW) coupled with a zero-order vortex half-wave retarder provided radially polarized excitation. For performance comparison, two objective lenses were employed: a 60 x magnification, 1.4 NA Nikon Plan Apochromatic oil immersion lens for high NA performance, and a 40 x magnification, 0.75 NA Nikon Plan Fluor dry lens for lower NA. Scattered light was collected using an Andor Shamrock SR 303i spectrometer with an iDus CCD camera, coupled via an f-number-matched free-space lens. We first demonstrate the PTTP`s capability to generate the localized enhancement of the excitation field and collect the scattered TERS signal. For this, we used a standard graphene sample mechanically exfoliated in a  170 $\mu$m thick glass coverslip. Then, we conducted two sets of TERS experiments on the same region of the graphene sample, first with the dry, 0.75 NA objective lens, and later with the oil immersion 1.4 NA objective lens. This order was important to preserve a dry coverslip for the dry objective lens.

\section{Results}
\begin{figure}[htbp]
\centering
\includegraphics[width=8.4cm]{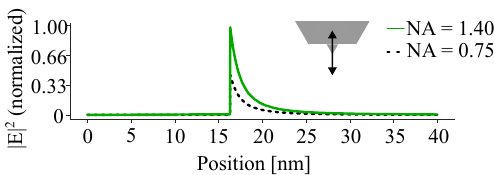}
\caption{$|$E$|^2$ sampled along the optical axis (see double arrow in inset) for NA = 0.75 (black-dashed) and NA = 1.4 (green-solid) configurations. Data normalized by the maximum amplitude of $|$E$|^2$ for NA = 1.40.}
\label{alongaxis}
\end{figure}

Figure \ref{alongaxis} illustrates the simulated electric field intensity along the optical axis of the TERS tip (see double-arrow in the inset), a PTTP structure with L = 420 nm and W = 1.2 $\mu$m. The simulations are performed in two excitation field scenarios, equivalent to the illumination using a NA = 1.4 (green-solid curve) and a NA = 0.75 (black-dashed curve). Notice that the NA = 0.75 illumination produces an enhancement at the tip`s apex that is approximately half of that of the NA = 1.4 configuration. This is unexpected when considering that the longitudinal field component of a tightly focused radially polarized field, i.e., the component along the tip axis, exhibits a steep dependence on the numerical aperture. Theoretically, when changing from NA = 1.4 to NA = 0.75, the field intensity component along the tip axis is expected to reduce at least by a factor of 6 \cite{novotny2012principles}. Our simulations indicate that the energy from the excitation laser is not transferred to a dipole-like structure, but to the bi-cavity plasmonic nanostructures (plateau + pyramid), with a higher efficiency in energy collection and localization in the tip apex.

\begin{figure}[htbp]
\centering
\includegraphics[width=8.4cm]{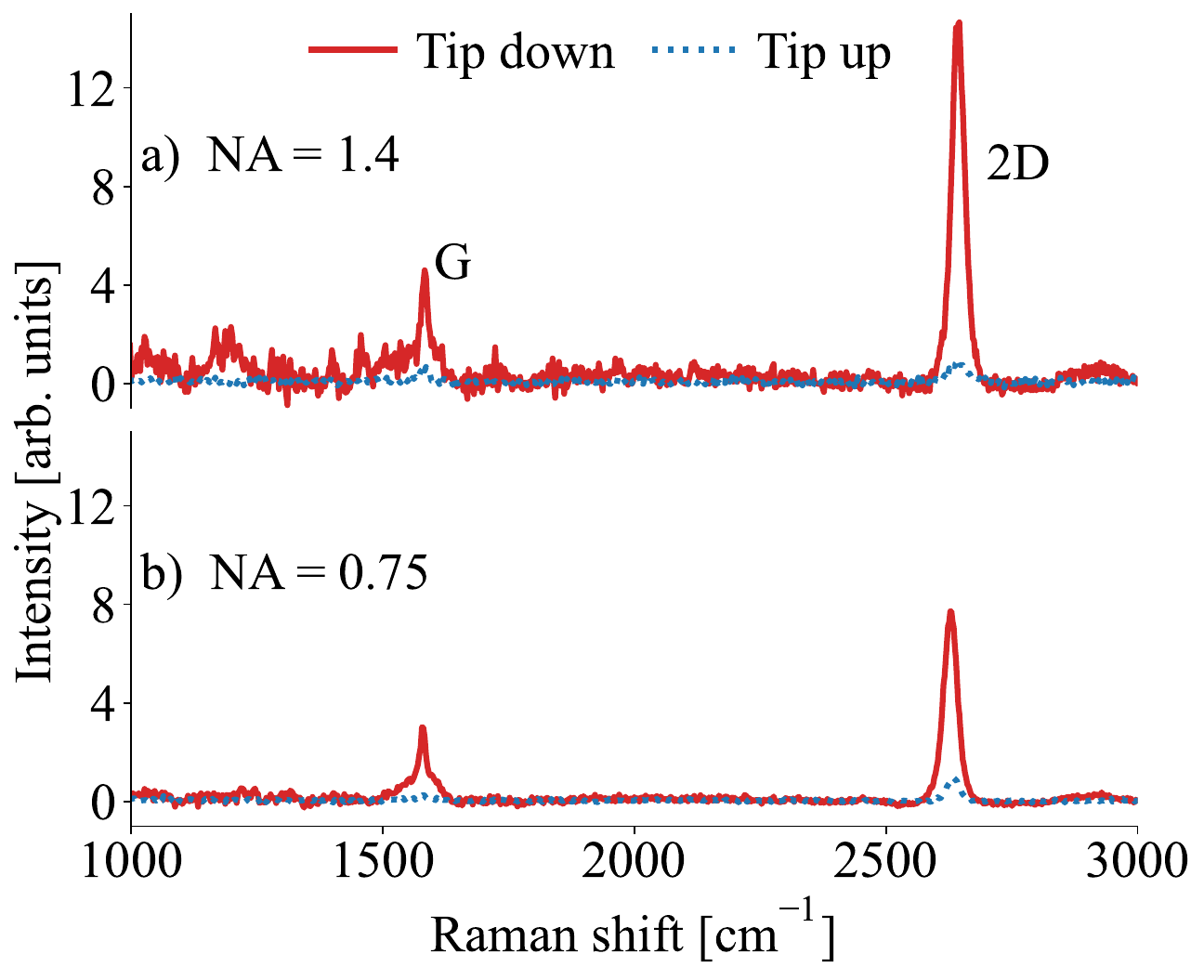}
\caption{Tip up (FF, in blue-dashed) and tip down (NF, in red-solid) Raman spectra measured from a monolayer graphene with the two illumination configurations: NA = 1.4 in a) and NA = 0.75 in b). All spectra were normalized by the amplitude of the 2D peak in the tip-up configuration, and the background was removed.}
\label{spectra}
\end{figure}

To confirm the simulated results, we conducted comparative experiments using objective lenses with different numerical apertures (NA) to assess their impact on TERS performance and the overall optical configuration. Figure \ref{spectra} presents two pairs of spectra, each consisting of a far-field (FF, in blue) and a near-field (NF, in red) spectrum, measured at a graphene monolayer and with different objective lenses (NA = 0.75 and NA = 1.4). This confirmed the efficiency of the bi-cavity plasmonic nanostructure to efficiently collect and emit propagating light in the TERS experiment.

\begin{figure}[h!]
\centering
\includegraphics[width=8.4cm]{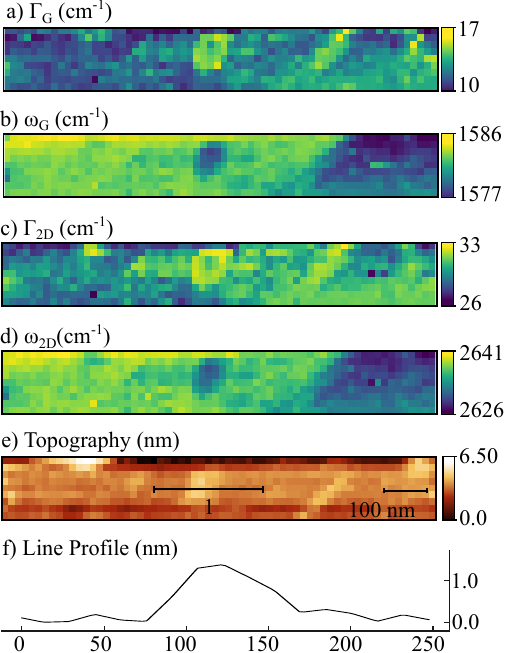}
\caption{Tip Enhanced Raman Hyperspectral maps of graphene measured with the 0.75 NA dry objective lens.
a,c) Full width at half maximum ($\Gamma$) and b,d) position ($\omega$) of the G and 2D bands, respectively. 
e) Atomic force microscopy (AFM) topography map. f) Topography line profile along segment 1 shown in e) of the feature responsible for the localized strain.}
\label{maps}
\end{figure}

To demonstrate that the low-NA result is efficient not only in receiving and emitting light in the FF, but also in generating localized information in the NF, using the 0.75 NA lens, we generated a hyperspectral map of a graphene sheet, as shown in figure \ref{maps}, in a region where an underlying nanometric topographical feature induced localized strain. This strain resulted in a blue shift of the graphene Raman peak positions $(\omega_{G})$ and $(\omega_{2D})$ and broadening of the peak width $(\Gamma_{G})$ and $(\Gamma_{2D})$ as observed in \cite{Beams2015}. The high spatial resolution is demonstrated by the observation of the characteristic annular point spread function of TERS in graphene, as previously reported in \cite{Miranda2023}, using here the low NA objective lens. 

\section{Discussion}
\begin{figure}
\centering
\includegraphics[width=8.4cm]{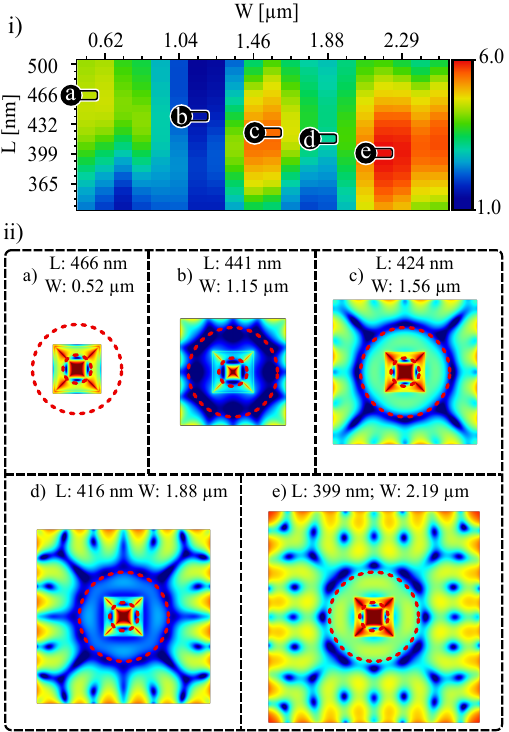}
\caption{i) Optimization heatmap revealing the field intensity($|$E$|$$^2$, color code) 5 nm below the tip apex for different tip configurations, varying the nanopiramid length (L) and plateau length (W). a-e locate specific configurations displayed in (ii). ii) from a) to e) Tangential $|$E$|$$^2$ is displayed for each point of interest along the surface of the plateau and nanopyramid as shown in figure \ref{pttp} b) (i). The region between the concentric dashed red circles represents the half-maximum contour of the radial excitation of the nanostructure.}
\label{curve}
\end{figure}

We now explain why an antenna-cavity bi-structure enables efficient far-field collection and strong near-field localization for bottom-illumination TERS without relying on high-NA optics. In figure ~\ref{curve},  we present an optimization heat map obtained by sweeping the nanopyramid length \(L\) and the plateau length \(W\) under NA \(= 0.75\) radially polarized excitation (Fig.~\ref{curve}\,(i)). The color scale reports the apex-localized field enhancement, and for selected \((L,W)\) points marked (a–e) in Fig.~\ref{curve}\,(i), the tangential \(|E|^{2}\) was sampled along the plateau and pyramid surfaces to correlate hotspots with spatial distribution of the tangential electric field intensity heatmaps shown in Fig.~\ref{curve}\,(ii). The optimization map displays a locus of maximum enhancement that spans specific \((L,W)\) values, indicating that the finite plateau functions as an in-plane Fabry-Pérot-like cavity for surface plasmon polaritons. The lateral edges act as reflectors that set a wavelength- and geometry-dependent behaviour, establishing standing-waves whose positions shift periodically with \(W\) and are directly visible in the tangential \(|E|^{2}\) plots (Fig.~\ref{curve}\,(ii)). In this picture, the PTTP nanopyramid behaves as a longitudinal optical antenna while the plateau supplies a lateral surface-plasmon-polariton cavity. When these subsystems are co-tuned at the optical drive wavelength, the structure supports a hybridized antenna-cavity state that funnels energy into the apex. This antenna-cavity hybridization framework resembles prior nanoparticle-on-mirror \cite{Tserkezis2015}.

\section{Conclusion}
In summary, finite-element simulations and experiments on graphene demonstrate that a dry NA\,=\,0.75 objective yields robust apex enhancement, reproducible NF spectra, and hyperspectral imaging that resolves the characteristic annular PSF. The central advance is to identify the plateau length $W$ as a design-critical parameter: $W$ forms an in-plane SPP Fabry--Pérot-like cavity that hybridizes with the longitudinal antenna mode set by $L$. Co-tuning $(L,W)$ creates a locus of maxima that funnels energy to the apex under radially polarized, on-axis excitation. These claims hold for the studied excitation (He-Ne, 632.8\,nm), PTTP dimensions, and substrate configuration, and they bound the conditions under which low-NA operation remains effective. Practically, relaxing the NA requirement increases working distance and broadens compatibility with constrained environments (e.g., thick substrates, sealed cells, UHV windows), moving TERS toward routine, deployment-ready nano-Raman spectroscopy technique.
\section{Acknowledgements}
The authors thank  Coordenacao de Aperfeicoamento de Pessoal de Nivel Superior (CAPES); FAPEMIG: (APQ - 04852-23, APQ - 01860-22, RED - 00081-23); CNPq (421469/2023-4).

\section{Conflict of Interest}

The authors declare no conflict of interest.

\section{Keywords}
	Nanostructures \textbullet\ 
	Nanotechnology \textbullet\ 
	Surface plasmon resonance \textbullet\ 
	Raman spectroscopy \textbullet\ 
	Scanning probe microscopy
    
\bibliographystyle{unsrt}
\bibliography{Monken_et_at_2025}

\end{document}